\documentclass[twocolumn,
aps,nofootinbib,showpacs,showkeys,preprint
tightenlines,preprintnumbers,] {revtex4}

\usepackage{epsf,epsfig,subfigure,graphicx,amsmath,amssymb}
\usepackage{color}
\newcommand{\dis}[1]{\begin{equation}\begin{split}#1\end{split}\end{equation}}

\newcommand{\OVER}[1]{\,\overline{\hskip -0.5mm #1}}

\newcommand{\etal}{{\it et al.}}

\newcommand{\gev}{\,\textrm{GeV}}
\newcommand{\mev}{\,\textrm{MeV}}

\newcommand{\ie}{{\it i.e.~}}

\newcommand{\NDW}{{N_{\rm DW}}}

\newcommand{\UPQ}{{U(1)$_{\rm PQ}$}}

\begin{document}
\draft

\title{\Large\bf High scale inflation, model-independent string axion, and QCD axion with domain wall number one}

\author{  Jihn E.  Kim }
\affiliation{
 Department of Physics, Kyung Hee University, 26 Gyungheedaero, Dongdaemun-Gu, Seoul 130-701, Republic of Korea
}

\begin{abstract}
The recent BICEP2 result shows that the Universe once has gone through the
vacuum  with the GUT scale energy density. The implied high scale inflation nullifies the dilution idea of topological defects, strings and domain walls, of the axionic system. In particular, domain walls are disastrous if the domain wall problem with $\NDW\ge 2$ is present. We argue that the model-independent axion in string compactification with the anomalous U(1)$_{\rm ga}$ gauge symmetry resolves the domain wall problem naturally with a symmetry principle.

\end{abstract}


\maketitle

\section{Introduction}

The recent report \cite{BICEP2I} on a possible significant nonzero B-mode polarization has attracted a great deal of attention \cite{Freese14,Lyth14,KimTrans14,Gondolo14,Recent}. One of the important implications of this result is that the Universe once had vacuum energy density at the order of the grand unification (GUT) scale, $\sim (2\times 10^{16\,}\gev)^4$. Even though this report has to be proved being consistent with the previous Planck data on the tensor to scalar ratio $r$ \cite{Planck13}, just the existence of a GUT scale vacuum energy density constrains many theoretical ideas suggested so far. In particular, this {\it high scale inflation} answers to a long-standing question in axion cosmology, ``Has the Peccei-Quinn (PQ) symmetry breaking occurred before or after the inflationary epoch?"

The hottest current issue \cite{Lyth14} is the reported large $r$ near $0.16$  \cite{BICEP2I} for which a trans-Planckian vacuum expectation value (VEV) of inflaton is needed for a large $e$-folding \cite{Lyth97}. Models leading to a trans-Planckian inflaton VEV are the Kim-Nilles-Peloso model  \cite{KNP04} and N-flation \cite{Nflation08}, which belong to a category of natural inflation \cite{Freese90}. Actually, another very interesting method from a string compactification view has been suggested recently for a trans-Planckian VEV from a discrete symmetry principle \cite{KimTrans14}.

Next hot issue is the {\it high scale inflation}. It seems that the PQ phase transition has occurred after (or at least at the end phase of) the inflation, and a possible scenario of the axion cosmology has been already scrutinized if cold dark matter (CDM) of the Universe  is 100\,\% axion \cite{Gondolo14}. The issue of the topological objects in axion models, domain walls \cite{Zeldovich74} and strings, has been discussed for a long time \cite{SikivieDW82,DavisR85,ChoiKimDW85, BarrChoiKim,SikivieString,Shellard94, Kawasaki12} to compare with the CDM axion energy density arising from the misalignment mechanism \cite{BCMaxion}. It has been pointed out that the axion domain wall number $\NDW$ should be one so that the energy density of axion walls does not overclose the Universe \cite{SikivieDW82}.

\begin{figure}[!t]
\begin{center}
\includegraphics[width=0.85\linewidth]{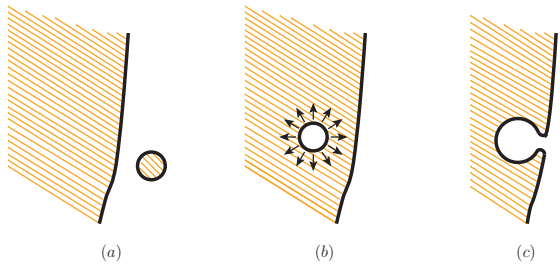}
\end{center}
\caption{Annihilation of cosmic scale string-wall system in the $\NDW=1$ model \cite{BarrChoiKim}.}\label{fig:DWOne}
\end{figure}

After axion gets a significant mass, the axion wall system is quickly erased if $\NDW=1$ as depicted in Fig. \ref{fig:DWOne}.
After axion strings are created, the axion string-wall system is assumed to have a scale invariant form \cite{Vilenkin82,BarrChoiKim}, and small scale walls surrounded by strings collide with the horizon scale string-wall system, effectively annihilating it. So, at present there is no energy crisis problem if $\NDW=1$. With  $\NDW=1$, the radiated axion spectrum from the string-wall system has been numerically estimated to give  ${\cal A}_\xi\sim 25$ \cite{Kawasaki12}, where ${\cal A}_\xi$ is the ratio of axions from radiation to the axions from the misalignment mechanism. On the other hand, the earlier estimate gave ${\cal A}_\xi\sim 1$ \cite{SikivieString}. Since Ref. \cite{Kawasaki12} did not take into account the efficient annihilation mechanism shown in Fig. \ref{fig:DWOne} above $T>400\,\mev$, we can take ${\cal A}_\xi\sim 5$ for an illustration as the geometrical mean value of Refs. \cite{Kawasaki12} and \cite{SikivieString}.

\begin{figure}[!t]
\begin{center}
\includegraphics[width=0.75\linewidth]{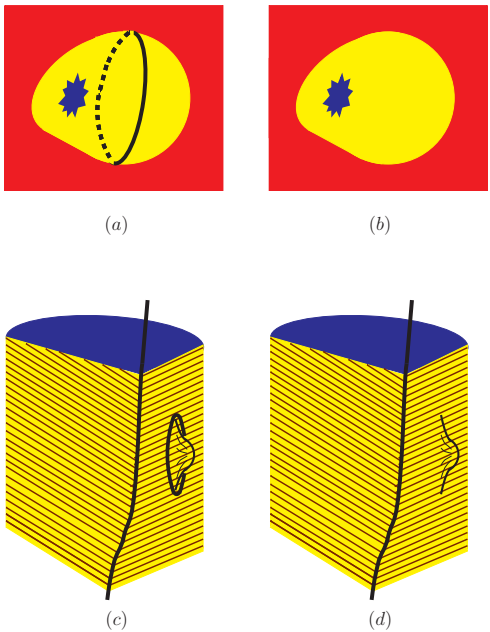}
\end{center}
\caption{Cosmic scale string-wall network system remaining in the $\NDW=2$ model. }\label{fig:DWTwo}
\end{figure}

On the other hand, if $\NDW\ge 2$, there is a severe energy crisis problem. In Fig. \ref{fig:DWTwo}, the string-wall system of $\NDW=2$ is shown. There are walls with strings and walls without strings as shown in (a) and (b), respectively. These small bubbles will collide with the horizon scale string-wall system, but the effect is not erasing the horizon scale wall as shown in (c) and (d) of Fig. \ref{fig:DWTwo}.

Another issue related to the PQ phase transition after inflation is the feasibility of QCD axion detection \cite{AxDet}. Our Solar system may belong to any value of $\theta_1$ from 0 to $\pi$, but the initial misalignment angle $\theta_1$ has the root mean square value  $\pi/\sqrt3$ if the PQ phase transition occurred after inflation \cite{Turner86}.
For this large $\theta_1$, the effect of anharmonic term is significant \cite{BaeHuhKim}. If ${\cal A}_\xi\simeq 5$, we have $f_a\lesssim (1.0-1.7)\times 10^{11\,}\gev$ \cite{Kawasaki12}. Even if  $\theta_1=\pi/\sqrt3$ is the most probable value we can expect, any other $\theta_1$ cannot be ruled out for the environment of Solar system since the small bubbles of different $\theta_1$ were possible after inflation. Therefore, it is necessary to look for the entire range of the allowed axion window.

Above all, the most important irreducible prediction in axion cosmology from this high scale inflation is that the axionic domain wall number should be one. As an example for $\NDW=1$ in the bottom-up approach, typically a KSVZ axion \cite{KSVZ} with one heavy quark is suggested. But, now we are discussing the particle spectrum at the end phase of inflation, around $10^{14\,}\gev$, and hence it is legitimate to bring out all the GUT scale spectrum.
The axionic domain wall number may not be just a DFSZ model \cite{DFSZ} with $\NDW=6$ or a KSVZ model with $\NDW=1$. We must add all contributions of the nonvanishing PQ charges \cite{KimRMP10}. There is one example of a top-down axion model with all the PQ chages of the quarks listed \cite{Kim88}.

In string models, string axions reside in the antisymmetric tensor field $B_{MN}\, (M,N=1,2,\cdots,10)$ among which $B_{\mu\nu}\,(\mu,\nu=1,\cdots,4)$ is the model-independent (MI) axion \cite{Witten84} and the rest are the model-dependent (MD) axions \cite{Witten85}. Both MI and MD axions are known to have their decay constants above the GUT scale \cite{ChoiKim85,Svrcek06}. In particular, the MI axion has the domain wall number one, which was used in  \cite{Kim88} to obtain $\NDW=1$ from a seemingly huge number of domain walls of order O(100) in string compactification. In view of the BICEP2 result, we revisit this solution of $\NDW=1$ QCD axion from the top-down approach.

\section{The effect of model-independent axion to QCD axion}

\begin{table}[!t]
\begin{center}
\begin{tabular}{|c|c|c|c||c  |}
\hline &&&  \\[-1.26em]
Fields&  U(1)$_{\rm ga}$ &  U(1)$_{\rm gl}$ & radial,\,phase&U(1)$_\Gamma$ \\
\hline\hline
$S_1$ &$1$ &$1$ &$\rho_1,\,\sigma_1$&  \\[0.3em]
$S_2$ &$1$ &$1$ &$\rho_2,\,\sigma_2$ & $1$ \\[0.3em]
 $\OVER{Q}_2$ &$1$ &$a$ &  & $-\frac12$  \\[0.3em]
$Q_2$ &$-2$ &$-1-a$ &  & $-\frac12$ \\[0.3em]
$\OVER{Q}_3$ &$1$ &$b$ & & $-\frac12$  \\[0.3em]
$Q_3$ &$-2$ &$-1-b$ & & $-\frac12$  \\[0.3em]
$\OVER{Q}_4$ &$1$ &$c$ & & $-\frac12$  \\[0.3em]
$Q_4$ &$-2$ &$-1-c$ & & $-\frac12$  \\[0.3em]
\hline
\end{tabular}
\end{center}
\caption{Three heavy quarks and two scalars for an illustration. In the second and third columns, the gauge and global U(1) charges are shown. For the scalar fields, the radial and phase fields are denoted as $\rho_i$ and $\sigma_i$, respectively. }\label{tab:Model}
\end{table}

For the axionic domain wall number in string compactification, one needs all information on the PQ charges of quarks, including the heavy quarks \cite{Kim88}. Instead, here we discuss the key issue in a field theory model with three heavy quarks as an easy example.
Below the scale $f_1$, we have fields with the gauge and global charges shown in Table \ref{tab:Model}. To mimic the Green-Schwarz term \cite{GS84}, let us introduce the nonrenormalizable coupling,
\dis{
\frac{g_3^2}{32\pi^2 f_1}\, \sigma_1 \, G\tilde G
}
where $G\tilde G$ is the anomalous combination of gluon fields $(1/2)\,\epsilon_{\mu\nu\rho\sigma} G^{\mu\nu} G^{\rho\sigma}$, and $\sigma_1$ corresponds to the MI axion whose decay constant $f_1$ is above the GUT scale \cite{ChoiKim85}. The Yukawa couplings respecting the gauge and global U(1) symmetries are
\dis{
{\cal L}_Y= \sum_{i=2}^4 \lambda_i S_2\OVER{Q}_i Q_i\,.\label{eq:Yukawa}
}
To mimic the Higgs mechanism for the anomalous U(1)$_{\rm ga}$ via the MI axion in string models, let us assign a vacuum expectation value $\tilde{V}$ to $S_1$,
\dis{
\langle S_1\rangle=\frac{\tilde{V}}{\sqrt2},
}
rendering the U(1)$_{\rm ga}$ gauge boson mass of $M_A=g\tilde{V}/2$. Below the scale $M_A$, we consider the fields $S_2$ and $Q_i\,(i=2,3,4)$, and the Yukawa couplings of Eq. (\ref{eq:Yukawa}). When both a gauge symmetry and a  global U(1) symmetry are broken by a single VEV, then there remains a global symmetry below the new gauge boson mass scale, which is called 't Hooft mechanism \cite{Hooft71}. The new global U(1) is denoted as U(1)$_\Gamma$. For the fields we consider, the charges $\Gamma$ are listed in Table \ref{tab:Model}.

The domain wall number calculated below $M_A$ looks like 3 because there are three heavy quark flavors,
\dis{
\frac{ g_3^2\, \sigma_2}{32\pi^2 \langle\rho_2\rangle} \,3 \, G\tilde G.
}
However, these three vacua are connected by the original anomalous U(1)$_{\rm ga}$ symmetry, and the physical domain wall number is 1. This identification of vacua by the Goldstone boson direction  can be called the Choi-Kim (CK) mechanism \cite{ChoiKimDW85}. The Lazarides-Shafi (LS) mechanism \cite{LS82,Kimprp87} where the vacua are identified by the center of nonabelian gauge group is not used here.\footnote{
Both for the CK and LS mechanisms, we imply only the discrete subgroups of global or gauge groups which commute with \UPQ. In the remainder of the paper, this is always implied.
}
Another idea breaking the degeneracy explicitly by the instanton effects of another nonabelian group \cite{BarrDW82} cannot automatically set $\theta_{\rm QCD}=0$ as the minimum, and we do not consider this possibility. The LS mechanism needs the invisible axion being housed in some nontrivial representation of the same gauge group  \cite{LS82}. For our model of Table \ref{tab:Model}, the singlet $S_1$ does not belong to SU(3)$_c$ and the center of SU(3)$_c$ cannot be used for the LS mechanism. The CK mechanism was noticed more than 25 years ago that if there are two axion directions $N_1$ and $N_2$, then the common divisor of $N_1$ and $N_2$ is the physically distinguishable domain wall number  \cite{ChoiKimDW85}. Our MI axion $\sigma_1$ connects the different axion vacua.

\begin{figure}[!t]
\begin{center}
\includegraphics[width=0.85\linewidth]{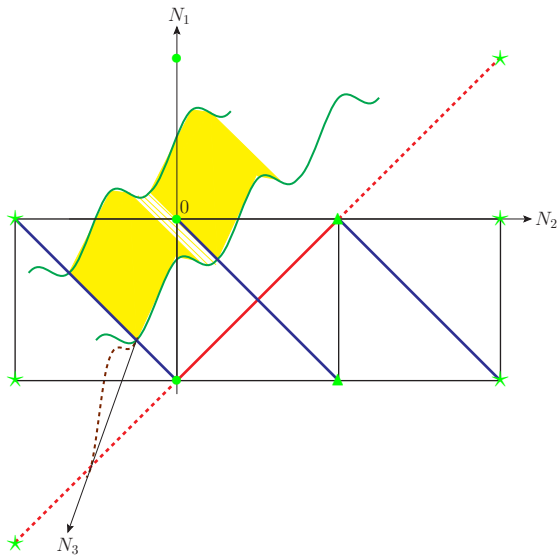}
\end{center}
\caption{A schematic view of physical domain wall number \cite{ChoiKimDW85}. The QCD axion direction is colored red along which the axion potential height is shown as the yellow band, and the flat valley  is colored blue. The physical domain wall number is 1 because three valleys are in fact connected. With an additional confining force, the additional axion direction is shown as $N_3$. The dashed brown curve depicts the mountain ridge of the hidden sector axion.}\label{fig:NeqOne}
\end{figure}

The CK idea is shown in Fig. \ref{fig:NeqOne} first for $N_1=1$ and $N_2=3$ with one confining force in the $N_2$ direction \cite{ChoiKimDW85,Kimprp87}. The case with two confining forces will be discussed in the next section.  The $N_1$ torus identifies the vacua with star marks, the vacua with triangle marks and vacua with bullets in the vertical direction. The $N_2$ torus identifies the vacua with stars in the horizontal direction. Therefore, the star, triangle and bullet vacua are identical, and we interpret it as the seemingly three $N_2$ domain walls are connected by the $N_1=1$ Goldstone boson direction, and hence there is only one physical domain wall. It is because the largest common divisor of $N_1$ and $N_2$ is 1. In Fig. \ref{fig:NeqOne}, the axion direction is colored red and the flat valley is colored blue.

The $N_1\equiv N_{\rm MI}=1$ role in the QCD axion has been noted earlier in string compactification models with an anomalous U(1)$_{\rm ga}$  gauge symmetry \cite{Kim88}, except which there has not appeared any QCD axion study in string models related to $N_{\rm MI}=1$. Our PQ symmetry is not an approximate one envisioned in Refs. \cite{KimPLB13, KimNilles14,KimTrans14} because the original anomalous U(1)$_{\rm ga}$  is the string-allowed gauge symmetry.  The resulting PQ symmetry is exact except for the anomalies of  \UPQ-SU(3)$_c$-SU(3)$_c$ and  \UPQ-SU(2)$_W$-SU(2)$_W$.

 In Fig. \ref{fig:NeqOne}, one can close the blue lines and observe that the VEV of the radial field  is three times larger than the shift along the one unit solid line. Thus, the scalar VEV is the original vacuum degeneracy times larger than $f_a$. Typically, in string compactification this vacuum degeneracy is expected to be of order 100. In \cite{Kim88}, the degeneracy is obtained to be 120. So, if $f_a\sim 10^{11\,}\gev$, the scalar VEV can be of order $10^{13\,}\gev$.

\section{Two confining forces and MI axion}

In supergravity models, a SUSY breaking sector is needed. The most plausible SUSY breaking sector is another confining force, a hidden sector nonabelian gauge group for which SU($N_h$) is assumed here \cite{Hidden}. To set the vacuum angles at zero ($\theta_{\rm QCD}=0$ and $\theta_{h}=0$), then we need two axions whose directions are denoted as $N_2$ (for SU(3)$_c$) and $N_3$  (for SU($N_h$)) in  Fig. \ref{fig:NeqOne}.   The direction of the MI axion is denoted again as $N_1\equiv N_{\rm MI}$. The anomaly coupling of the MI axion to these nonabelian gauge bosons is
\dis{
\frac{a_{MI}}{32\pi^2 F_{MI}}\left( G\tilde G+ F_h\tilde F_h \right)
}
where $F_h\tilde F_h$ is the hidden sector anomaly term and gauge couplings are absorbed in the field strengths. The MI axion has the same coefficient for the anomaly couplings \cite{Witten84}. Also, the sum of the anomalous U(1)$_{\rm ga}$ charges are the same for all gauge groups obtained from the heterotic string \cite{Kim88,Munoz89}; thus the QCD axion $a_2$ and the hidden sector axion $a_3$ couple to the respective nonabelian group gauge bosons as
\dis{
\frac{ {\cal N}a_{2}}{32\pi^2 f_{2}}  G\tilde G
+\frac{ {\cal N}a_{3}}{32\pi^2 f_{3}} F_h\tilde F_h,
}
where ${\cal N}$ is of order 100.
We can consider three orthogonal axion currents, applicable to the pseudoscalar particles using the PCAC relations,
\dis{
&\partial_\mu \theta_e \propto \frac{\partial_\mu a_2}{f_2} -\frac{\partial_\mu a_3}{f_3},\\
&\partial_\mu \theta_f\propto \frac{\partial_\mu a_{MI}}{F_{MI}}+ {\cal N} \left(\frac{\partial_\mu a_2}{f_2} +\frac{\partial_\mu a_3}{f_3}\right) ,\\
&\partial_\mu \theta_g\propto {\cal N}\frac{\partial_\mu a_{MI}}{F_{MI}}- \left(\frac{\partial_\mu a_2}{f_2} +\frac{\partial_\mu a_3}{f_3}\right) .
}
Note that along the $\theta_g$ direction, there is no anomaly coupling, viz. $\partial^2\theta_g=0$, and  the $\theta_g$ direction is the flat direction shown as the blue valley in  Fig. \ref{fig:NeqOne}.$^1$ It is tantalizing to notice that string compactification with the anomalous U(1)$_{\rm ga}$ gauge symmetry leads to the QCD axion with $\NDW=1$.

Having established that the string compactification with the anomalous U(1)$_{\rm ga}$  gauge symmetry guarantees $\NDW=1$ both for the hidden-sector and the QCD-sector anomalies, we can glimpse the Yukawa coupling  of the QCD axion and electron. The DFSZ axion coupling to electron is 6 times weaker than $1/f_a$ because the $\NDW=6$ in the DFSZ model, but the QCD axion originating from string along the above scenario is about 100 times weaker than $1/f_a$, making it more difficult to probe the existence of axion--electron coupling.

If both the confining scales and the decay constants are widely separated, there exists a cross theorem that the larger confining scale chooses the smaller decay constant \cite{KimJHEP99}. Since the hidden sector is expected to have the higher anomaly potential, the smallest decay constant corresponds to the hidden sector axion. The QCD axion potential chooses the next to the smallest decay constant. Both of these QCD and hidden-sector axion decay constants are expected to be at the intermediate scale. The lowest height chooses the largest decay constant, \ie in the present case the MI axion potential is flat and the MI axion chooses $f_{\rm MI}\,(>10^{16\,}\gev$).

\section{Conclusion}
The recent BICEP2 result implies that the inflationary phase has ended at $H_I\simeq 10^{14\,}\gev$ which is above the expected QCD axion window: $f_{a,\gev}=[10^{10},10^{12}]$. This requires that the axion domain wall number $\NDW$ should be one. We have shown that the MI axion in string compactification with the anomalous U(1)$_{\rm ga}$  gauge symmetry connects different domains, which realizes a long-sought  natural solution (from a symmetry argument) for  $\NDW=1$ models. Even if the final value of the tensor to scalar ratio $r$ is reduced by a factor of few, our conclusion requiring $\NDW=1$  is not changed because the fact that the PQ phase transition has occurred after inflation would not be changed.

\section*{Acknowledgments}
This work is supported in part by the National Research Foundation (NRF) grant funded by the Korean Government (MEST) (No. 2005-0093841) and  by the IBS(IBS CA1310).


\end{document}